\documentclass[aps,pra,superscriptaddress,11pt]{revtex4-2}
\usepackage{amsfonts}
\usepackage{amssymb}
\usepackage{amsmath}
\usepackage{booktabs}
\usepackage{graphicx}
\usepackage{dcolumn}
\usepackage{bm}
\usepackage{units}
\usepackage{multirow}
\usepackage{CJKutf8}
\usepackage{subfigure}
\usepackage{color}
\usepackage{xcolor}
\usepackage{url}
\usepackage[colorlinks,linkcolor=blue,anchorcolor=blue,citecolor=blue,urlcolor=blue]{hyperref}

\begin{document}

\title{Observation of Anderson localization transitions in a two-dimensional conjugated metal-organic framework}

\author{Jinhao Cheng}
\affiliation{Key Laboratory of Organic Integrated Circuits, Ministry of Education \& Tianjin Key Laboratory of Molecular Optoelectronic Sciences, Department of Chemistry, School of Science, Tianjin University, Tianjin, 300072, P. R. China.}
\affiliation{Collaborative Innovation Center of Chemical Science and Engineering (Tianjin), Tianjin 300072, China}
\author{Chen Wang}
\affiliation{Center for Joint Quantum Studies and Department of Physics, School of Science, Tianjin University, Tianjin 300350, China}
\author{Wenxue He}
\affiliation{Center for Joint Quantum Studies and Department of Physics, School of Science, Tianjin University, Tianjin 300350, China}
\affiliation{Tianjin Key Laboratory of Low Dimensional Materials Physics and Preparing Technology, School of Science, Tianjin University, Tianjin 300072}
\author{Jiaojiao Wang}
\affiliation{Center for Joint Quantum Studies and Department of Physics, School of Science, Tianjin University, Tianjin 300350, China}
\affiliation{Tianjin Key Laboratory of Low Dimensional Materials Physics and Preparing Technology, School of Science, Tianjin University, Tianjin 300072}
\author{Yifan Pang}
\affiliation{Key Laboratory of Organic Integrated Circuits, Ministry of Education \& Tianjin Key Laboratory of Molecular Optoelectronic Sciences, Department of Chemistry, School of Science, Tianjin University, Tianjin, 300072, P. R. China.}
\affiliation{Collaborative Innovation Center of Chemical Science and Engineering (Tianjin), Tianjin 300072, China}
\author{Fan Yang}
\affiliation{Center for Joint Quantum Studies and Department of Physics, School of Science, Tianjin University, Tianjin 300350, China}
\affiliation{Tianjin Key Laboratory of Low Dimensional Materials Physics and Preparing Technology, School of Science, Tianjin University, Tianjin 300072}
\author{Shuaishuai Ding}
\email[Corresponding author: ]{dingshuaishuai@tju.edu.cn}
\affiliation{Collaborative Innovation Center of Chemical Science and Engineering (Tianjin), Tianjin 300072, China}
\affiliation{Key Laboratory of Organic Integrated Circuits, Ministry of Education \& Tianjin Key Laboratory of Molecular Optoelectronic Sciences, Department of Chemistry,  Institute of Molecular Aggregation Science, Tianjin University, Tianjin, 300072, P. R. China.}
\author{Hechen Ren}
\email[Corresponding author: ]{ren@tju.edu.cn}
\affiliation{Center for Joint Quantum Studies and Department of Physics, School of Science, Tianjin University, Tianjin 300350, China}
\affiliation{Tianjin Key Laboratory of Low Dimensional Materials Physics and Preparing Technology, School of Science, Tianjin University, Tianjin 300072}
\affiliation{Joint School of National University of Singapore and Tianjin University, International Campus of Tianjin University, Binhai New City, Fuzhou 350207, China}
\author{Wenping Hu}
\email[Corresponding author: ]{huwp@tju.edu.cn}
\affiliation{Key Laboratory of Organic Integrated Circuits, Ministry of Education \& Tianjin Key Laboratory of Molecular Optoelectronic Sciences, Department of Chemistry, School of Science, Tianjin University, Tianjin, 300072, P. R. China.}
\affiliation{Collaborative Innovation Center of Chemical Science and Engineering (Tianjin), Tianjin 300072, China}
\affiliation{Joint School of National University of Singapore and Tianjin University, International Campus of Tianjin University, Binhai New City, Fuzhou 350207, China}


\begin{abstract}
Anderson localization transitions are a universal quantum phenomenon sensitive to the disorder and dimensionality of electronic systems. Over the past decades, this intriguing topic has inspired overwhelmingly more theoretical studies than experimental verifications due to the difficulty of controlling a material's disorder or dimensionality without modifying its fundamental electronic properties. Organic crystals with their rich disorders would be terrific playgrounds to investigate such disorder-driven phase transitions except for their low conductivities which usually prohibit low-temperature measurements. Here, we conduct systematic transport experiments in mesoscopic devices made with copper benzenehexathiol thin films across a wide range of thicknesses. We find metal-insulator transitions both among three-dimensional samples with different disorder strengths and between three-dimensional and quasi-two-dimensional samples. Temperature-dependence analysis of the conductivities corroborates the dimensionality crossover. Moreover, our theoretical modeling provides a basis for understanding both types of metal-insulator transitions within the framework of Anderson localization transitions. Our findings establish for the first time that organic crystals such as conductive metal-organic frameworks can exhibit such quantum interference effects. With organic materials’ versatile chemical designs and crystalline structures, our work opens new avenues to search for novel quantum phenomena in organic material platforms.

\end{abstract}
\maketitle
\newpage

Anderson localization transition (ALT), a disorder-driven quantum phase transition from extended to localized states, is a classic topic in condensed matter physics that receives continuous research spotlight\cite{kramer_localization_1993,lagendijk_fifty_2009}. A representative case is the absence of diffusion of an initially localized wave packet at arbitrarily weak disorders in one dimension where no transition occurs. On the other hand, ALTs can occur in three dimensions, manifested in transport measurement as a disorder-driven metal-insulator transition. Such transitions have been observed in experimental systems, e.g., the doped semiconductors\cite{raj_angle-resolved_2006,ying_anderson_2016} and the persistent photoconductors\cite{pirralho_transition_2017}.
\par

In the intermediate regime, two-dimensional (2D) electronic systems always support localized states and do not exhibit the ALT, except maybe in rare cases with a strong delocalization effect from the spin-orbit coupling\cite{hikami_spin-orbit_1980}. The unique scenario of investigating the transition from a three-dimensional (3D) disordered system to its 2D counterpart presents a complex and challenging task. The strict two-dimensionality of inorganic crystals, often requiring a thickness on the order of nanometers or below, introduces complexities such as altered band structures and the influence of surface states on observed transport phenomena. Furthermore, the introduction of disorders in inorganic crystals, typically through extrinsic dopants, adds another layer of complexity beyond a random potential distribution in the crystal lattice\cite{ying_anderson_2016}. 
\par

On the other hand, organic crystals come with rich possessions of disorders, making them ideal testbeds to study this problem. However, due to the low conductivities of organic crystals, ALTs have not been studied in organic electronic systems via transport measurement. Some above-room-temperature metal-insulator-like transitions have been reported in a biological metal-organic framework (MOF) generated upon coordination of the cysteine and the cupric ion, but the phenomenon is rooted in a different nature\cite{sindhu_insulator--metal-like_2023}. 
\par

In recent years, MOFs have attracted research interest from multiple disciplines thanks to their chemical tunability, which leads to broad applications in gas storage\cite{connolly_tuning_2019,connolly_shaping_2020}, gas separation\cite{hiraide_high-throughput_2020,sun_oriented_2023}, catalysis\cite{shen_programmable_2021,yuan_large-area_2024}, and sensors\cite{zhang_ultrasensitive_2022,moumen_metal-organic_2022}. Among them, two-dimensional conjugated MOFs (2D-cMOFs) exhibit outstanding intrinsic conductivity thanks to their extended $\pi$-$d$ conjugation in the 2D plane\cite{sun_electrically_2016}. They have already been widely applied in the fields of batteries\cite{nam_conductive_2019,luo_design_2022}, optoelectronics\cite{he_confinement_2019,shang_one-dimensional_2022} and thermoelectrics\cite{erickson_thin_2015,lu_precise_2022,Hio_BHTTE_2024}. Local gate control of Mott metal-insulator transitions has also been demonstrated in a 2D-cMOF consisting of 9,10-dicyanoanthracene (DCA) molecules and copper atoms, highlighting the interesting many-body physics in this material family\cite{lowe_local_2024}. Copper benzenehexathiol ($\mathrm{Cu_{3}BHT}$) is a member of 2D-cMOFs with a record high electrical conductivity under room temperature\cite{huang_two-dimensional_2015}. The structure of $\mathrm{Cu_{3}BHT}$ is shown in Fig. 1a. Copper atoms are bridged by BHT molecules via Cu–S coordination bonds, forming a perfect Kagome lattice. The Fermi surface contour in the reciprocal space of single-layer $\mathrm{Cu_{3}BHT}$ is shown in Fig. 1b, where we can see several hole pockets at $\mathit{\Gamma}$ point, $K$ point, and $M$ point arising from $p$-orbitals of C and S atoms along with $d$-orbitals of Cu atoms\cite{zhang_theoretical_2017}. In the band structure of bulk $\mathrm{Cu_{3}BHT}$, bands crossing the Fermi level can also be observed, in agreement with the metallic state observed in 3D $\mathrm{Cu_{3}BHT}$ samples\cite{huang_superconductivity_2018, Gao_BulkCuBHT_2019}. The extraordinary conductivity and electronic band structures of the 2D-cMOF with Kagome lattice, making it a promising candidate for hosting exotic condensed-matter-physics phenomena such as unconventional superconductivity\cite{huang_superconductivity_2018,Pan_BHTSC_2024} and quantum spin liquid\cite{takenaka_strongly_2021,ZhangQichun_OrganicQuant_2023}. 
\par

In this work, we conduct systematic transport experiments in mesoscopic devices made with $\mathrm{Cu_{3}BHT}$ thin films and observe metal-insulator transitions induced by disorder strength and dimensionality crossover. By controlling the synthesis process, we obtain samples with different thicknesses and divide them into three groups: the quasi-2D group, the intermediate group, and the 3D group. By analyzing the temperature dependence of resistivity in both metallic and insulating samples, we investigate charge transport physics in different sample groups. For samples in the quasi-2D group, arbitrarily weak disorders can easily localize the electronic states (Fig. 1c). However, with the change of sample dimensions, samples with the same level of disorders can gradually change into a transitional state with less localized carriers (Fig. 1d) and finally allow a metallic state with delocalized carriers (Fig. 1e). Meanwhile, within the 3D group, increase in strength of disorder can lead to localized carrier in 3D samples (Fig. 1f). Using finite-size scaling analysis of participation ratios of a disordered tight-binding model, we can theoretically understand metal-insulator transitions within the framework of ALTs, where the electronic states lose mobility from disorder-induced quantum interference effects.
\par\quad\par

\noindent{\large\textbf{Thickness-controlled synthesis of $\mathrm{Cu_{3}BHT}$ thin films}}\\
$\mathrm{Cu_{3}BHT}$ films were synthesized by liquid-liquid interfacial reaction as shown in Fig. 1g. The process of crystal growth can be divided into three stages. The first stage is nucleation, in which copper ions coordinate with BHT molecules and form crystal nuclei on the interface. Then the coordination continues on the interface and the crystals grow horizontally due to the interface-confining effect. With a layer of $\mathrm{Cu_{3}BHT}$ covering the interface, the copper ions have to diffuse through the $\mathrm{Cu_{3}BHT}$ to coordinate with the BHT molecules, which is the thickening process. During the thickening process, the thickness of the film can be tuned by changing the precise concentration of copper ions since it can change the chemical potentials which drive the diffusion process\cite{liu_thermodynamic_2024}. We used different concentrations of copper ions in the water phase and obtained samples with thicknesses ranging from 30 nm to 1000 nm (Fig. 2a). After being thoroughly washed with degassed water and organic solutions, $\mathrm{Cu_{3}BHT}$ films were transferred onto different substrates for subsequent measurement.
\par

The composition and crystal structure of $\mathrm{Cu_{3}BHT}$ films are characterized via multiple techniques. Fourier transform infrared spectrometer (FT-IR) is applied to determine the bonding between metal ions and BHT. The vanished peak of the sulfur–hydrogen bond at 2490 $\mathrm{cm^{–1}}$ indicates that sulfur atoms of BHT molecules successfully coordinate with copper atoms (Fig. S1). X-ray photoelectron spectroscopy (XPS) is used to characterize the chemical composition of $\mathrm{Cu_{3}BHT}$ films, revealing the presence of copper, sulfur, and carbon in the film (Fig. S2a). High-resolution scanning of the Cu 2$p$ region indicates that copper atoms in samples with different thicknesses show Cu(I) and Cu(II) form at the same time. The phenomena has been previously reported in $\mathrm{Cu_{3}BHT}$ and other copper coordinated MOFs and attributed to different coordination geometry, indicating strong charge transfer between copper ions and BHT ligands which may contribute to the in-plane $\pi–d$ conjugation and conductivity\cite{huang_two-dimensional_2015,Wang_copperstate_2023, Hio_BHTTE_2024} (Fig. S2b). Energy dispersive spectrometer (EDS) mapping shows the even distribution of copper, sulfur, and carbon in $\mathrm{Cu_{3}BHT}$ films (Fig. S3). Transmission electron microscopy (TEM) images in Fig. S4a and atomic force microscope (AFM) images in Fig. 2b are taken to illustrate the layered structure of synthesized films. High-resolution transmission electron microscopy (HRTEM) images in Fig. S4b and Fig. S4c and powder X-ray diffraction (PXRD) patterns in Fig. 2c of films with different thicknesses match well with the structure of $\mathrm{Cu_{3}BHT}$ crystals\cite{huang_superconductivity_2018,huang_BHT_2020}. Selected area electron diffraction (SAED) patterns of the films show clear spots matching well with crystal indices of $\mathrm{Cu_{3}BHT}$, which reveal the high crystallinity of our films (Fig. S5).  All the above characterizations confirm the successful preparation of $\mathrm{Cu_{3}BHT}$ crystal.
\par

Defects and disorder have an important bearing on MOFs' transport properties\cite{Cheetam_Defects_2016}. XPS spectrum in Fig. S2 indicates that neither impurity atoms nor wrong bonding type which may induce structural disorder appeared in our sample. From the surface morphology pictures obtained by AFM, mesoscopic scale defects like cracks and pores are rarely seen except in samples thinner than 10 nm (Fig. S6, Fig. 2d and 2e). We also observe that the surface roughness of $\mathrm{Cu_{3}BHT}$ films has a positive correlation with the film thicknesses (Fig. 2f) and roughnesses of the top and bottom surfaces are different (Fig. S7) from AFM images. The phenomenon can be attributed to the weakened interface-confining effect as the films thicken and the reaction zone moves away from the liquid-liquid interface\cite{Rao_liquidinterface_2008}. PXRD patterns of samples with different thicknesses in Fig. 2c indicate that the crystallinity differs only slightly. 

HRTEM images in Fig. S4c also show disorders like amorphous region, grain boundaries and lattice mismatch in our samples. The domain sizes of $\mathrm{Cu_{3}BHT}$ films with different thicknesses are all on the order of 100 nm in Fig. S8. Kelvin probe force microscopy (KPFM) tests are carried out to investigate the charge distribution on the surface of $\mathrm{Cu_{3}BHT}$. Despite the granular nature, the $\mathrm{Cu_{3}BHT}$ crystal shows even surface charge distribution and high charge density (Fig. S9). The above spectrums and images help us clarify that the disorder in these thin films mainly originates from domain boundaries rather than other chemical, structural, or doping factors.
\par\quad\par

\noindent{\large\textbf{Room-temperature electrical properties of layered $\mathrm{Cu_{3}BHT}$}}\\
The layered structure and chemical composition of $\pi$-conjugated 2D coordination polymers typically lead to different charge transport pathways vertically and laterally and therefore different intra- and inter-layer conductivities\cite{wang_interfacial_2021}. For $\mathrm{Cu_{3}BHT}$, Cu-S coordination bonds and the benzene core of BHT can construct extended $\pi–d$-conjugated planes which contribute to in-plane charge transport, while $\pi–\pi$ stacking of benzene cores plays a crucial role in out-of-plane charge transport. 
\par

For room-temperature characterization of $\mathrm{Cu_{3}BHT}$'s intra- and inter-layer conductivities, we conducted probe-station measurements on samples transferred onto $\mathrm {Si/SiO_2}$ substrates with thermally evaporated parallel gold electrodes. Four-probe measurements of both 100-nm-thick and 800-nm-thick samples give lateral conductivities on the order of $10^{2}$ S/cm (Fig. 2g). To obtain the vertical conductivity, we deposited another group of parallel electrodes on top of the transferred sample crossing the bottom electrodes to create overlapping areas of 50 $\mu$m × 50 $\mu$m. The structure of device is shown in Fig. S10. By comparing results from four-probe and two-probe lateral configurations, we obtain the contact resistance between the sample and the electrodes and subtract it from the two-probe measurement of the vertical resistance. This gives vertical conductivities on the order of $10^{-3}$ S/cm for both 100-nm-thick and 800-nm-thick samples (Fig. 2h). These results suggest charge transport pathways exist both in-plane and out-of-plane with drastically different amplitudes, making $\mathrm{Cu_{3}BHT}$ a layered material regardless of its thickness. 

Complimentary to the probe-station method, tunneling AFM (TUNA) measurements provide local vertical conductivity information, revealing the microscopic domain structures. Surface morphology and electric current are simultaneously scanned by applying a bias of 5 V between the tip and the bottom electrodes of the sample. The domains and boundaries are distinguishable, and the lateral sizes of the domains are on the order of 100 nm (Fig. S11), consistent with our HRTEM results. Local current-voltage measurements within domains display linear relations, suggesting ohmic conductance inside domains (Fig. 2i). The estimated vertical conductivity using TUNA is on the order of $10^{-2}$ S/cm, which is larger than we get from the probe station, suggesting the latter to be a macroscopic average of inhomogeneous conductivities over domains and boundaries. The contact area between tip and sample is considered to be 100 $\mathrm{nm^{2}}$\cite{AFMtipSize}.
\par\quad\par

\noindent{\large\textbf{Transport properties and metal-insulator transitions of $\mathrm{Cu_{3}BHT}$}}\\
To conduct transport measurements, $\mathrm{Cu_{3}BHT}$ films were cleaned and transferred onto Hall bars pre-patterned on $\mathrm {Si/SiO_2}$ substrates. The separation between the longitudinal voltage probes is sub-10 $\mu$m, providing a mesoscopic length scale for our measurements (Fig. S12). We studied the temperature dependence of over ten $\mathrm{Cu_{3}BHT}$ devices with different film thicknesses and observed a spectrum of behaviors ranging from metallic to insulating as shown in Fig. S13. Based on the observed temperature-dependence of electrical conductivity, we divide the measured samples coarsely into three groups by thickness: the quasi-2D group with thicknesses below 200 nm, the intermediate group with thicknesses ranging from 200 to 600 nm, and the 3D group with thicknesses ranging from 600 to 1200 nm. The thicknesses $t$ and room temperature conductivities $\sigma$ of the representative samples in Fig. 3a are shown in Table 1. Details of other samples are shown in Table S1.

\begin{table}[h]
	\caption{Contents of Samples in Fig. 3a}
	\centering
	\setlength{\tabcolsep}{3mm}{ 
	\begin{tabular}{ccccc}
	\toprule [1pt]     
	Sample & S3 & S5 & S10 & S14\\ 
	\midrule [1pt]  
	$t$ (nm) & 100 & 250 & 800 & 750\\ 
	$\sigma$ (S/cm) & 176.1 & 831.8 & 556.6 & 482.9\\ 
	\bottomrule [1pt]     
	\end{tabular}}
	\label{tab:example}
\end{table}

\par

We observe distinct metallic and insulating behaviors among $\mathrm{Cu_{3}BHT}$ samples from the 3D group. The conductivity of the metallic samples (represented by Sample S14) increases with decreasing temperature. Moreover, the exponent of its dependence on temperature changes from high- to low-temperature ranges, signaling multiple scattering mechanisms at play such as electron-phonon and electron-electron interactions typical in metallic conductors\cite{ziman_electronphonon_1963}. On the other hand, the conductivity of the insulating samples (represented by Sample S10) decreases with decreasing temperature. While its temperature dependence is inconsistent with a thermally activated gap in the form of $\sigma \propto  e^{-E_a/k_B T}$ (orange line in Fig. 3b), it agrees well with Mott's variable-range hopping (VRH) model where $ \sigma \propto e^{- (1/T) ^ \frac 1 {1+d}}$ with $d = 3$ here being the dimension of the materials (bottom panel of Fig. 3c)\cite{mott_conduction_1968,stallinga_electronic_2011}. This confirms the three-dimensionality of the insulating samples in this group and the granular origin of their scattering mechanism. Such non-linear relationships between the logarithm of the conductivity and inverse temperature is common among conducting polymers and coordination polymers, which points to a disorder-limited charge transport mechanism\cite{givaja_electrical_2011}. We hence attribute the difference between these contrasting metallic and insulating behaviors in the same thickness group to the inhomogeneity of disorder strengths among 3D $\mathrm{Cu_{3}BHT}$ samples, giving rise to a metal-insulator transition governed by classic Anderson localization. 
\par

Among the thinnest samples, the temperature dependence of electrical conductivity from the quasi-2D group (represented by Sample S3) exhibits exclusively insulating behaviors. Even though the room-temperature conductivity remains on the order of $10^2$ S/cm, the longitudinal conductivity rapidly drops as temperature lowers. Again, this conductivity-temperature relation is inconsistent with any thermally activated gap (brown line in Fig. 3b), indicating the insulating behavior in these thinnest samples does not originate from a bandgap opening. This is consistent with previous calculations, which confirms the metallic nature in $\mathrm{Cu_{3}BHT}$ even down to the single-layer limit, where multiple bands cross the Fermi level\cite{zhang_theoretical_2017}. We again find Mott's VRH model a good fit for the $\sigma$-$T$ curves of these samples, but this time with $d = 2$ being the dimension of the materials. The high $R^2$ coefficient of determination verifies the quasi-two-dimensionality of these thinnest samples (upper panel of Fig. 3c). We will explore this dimensionality-induced metal-insulator transition in more detail in the next section.
\par

The conductivity of samples from the intermediate group (represented by Sample S5) shows a similar trend in temperature to the quasi-2D group. However, the rate at which the system becomes insulating is much lower. The conductivity of the sample only decreases to half of room-temperature conductivity at 2 K, allowing transport measurements at low temperatures. By measuring the Hall effect at different temperatures, we extract the mobilities and carrier densities of a 600-nm-thick sample S8 from the intermediate group. Although the conductivity decreases at low temperatures, magnetoconductance and Hall voltage can still be observed in an external field (Fig. S14). The sign of the Hall voltage confirms that the dominant carrier type is holes, in agreement with previous work\cite{Nishihara_BHTcrystallinity_2022}. The calculated mobility and carrier density at 5 K are 4.82 cm$^{2}$·V$^{-1}$·s$^{-1}$ and 1.24$\times10^{12}$ $\mathrm{cm^{-2}}$ (Fig. 3f).
\par

Interestingly, the $\sigma$-$T$ curve of sample S5 in the intermediate group fits neither the thermal-activation model (green line in Fig. 3b) nor the VRH models for $d = 2$ or 3 (Fig. 3d). We can express the temperature dependence in the form of modified Drude conductivity, $\sigma$ =  $\sigma_D$ + $\delta \sigma_C$, where $\sigma_D$ is the temperature-independent Drude conductivity of metallic domains and $\delta \sigma_C$ is the temperature-dependent correction term\cite{Anderson_ScalingTheory_1979, Lee_DisorderdSystem_1985}. We find that the conductivity correction of Sample S5 in the intermediate group shows a temperature dependence of $\delta\sigma_C \sim T^{\frac 1 2}$ (Fig. 3e), which was observed in a previous experiment on $\mathrm{SrTi_{1-x}Ru_{x}O_3}$ and attributed to weak localization effect during a metal-insulator transition\cite{Kim_STROMIT_2005}.
\par

Lastly, we perform magneto-transport measurements in our mesoscopic $\mathrm{Cu_{3}BHT}$ devices by applying a magnetic field in the direction perpendicular to the sample plane. Most samples conductive at low temperatures show positive parabolic magnetoresistance. Yet, in two samples from the intermediate group, we observe a sharp resistance dip near zero field in addition to a parabolic magnetoresistance background, and the feature is repeated at various speeds and directions of the magnetic field sweep (Fig. S15). Given the potential presence of spin-orbit coupling in this material, we analyzed the data according to weak anti-localization theory, though the fitting result wasn’t satisfactory for either 3D or 2D regimes (see Supplementary Text S1). This suggests the resistance dip may have other origins such as partial superconductivity in small pockets of the sample. Given the temperature range we conducted experiments at was much higher than any previously reported $T_c$ of $\mathrm{Cu_{3}BHT}$, further studies are merited to verify a potential higher $T_c$ in thinner samples\cite{zhang_theoretical_2017}.
\par\quad\par

\noindent{\large\textbf{Theoretical insights on Anderson localization transitions}}\\
To obtain more insights into the observed metal-insulator transitions in $\mathrm{Cu_{3}BHT}$ films, we numerically investigate the localization nature of a disordered tight-binding model without losing generality. The Hamiltonian of our model reads
\begin{equation}
\begin{gathered}
H=\sum\limits_i{\epsilon_i}{c_i^{\dag}}{c_i}+t{\sum\limits_{\langle ij\rangle}}{c_i^{\dag}}{c_j}+h.c..
\end{gathered}\label{eq_2_2}
\end{equation}
\par

Here, $c_i^\dag$ and $c_i$ represent the single-particle creation and annihilation on a cubic lattice site $i=(n_x a,n_y a,n_z a)$ with $n_{x,y,z}$ being integers and $1\le{n_{x,y}}\le L_\parallel$,$1\le{n_z}\le{L_\perp}$ and $a=1$ being the lattice constant. $t=1$ is the hopping energy, and $\epsilon_i$ is the on-site energy distributing randomly and uniformly in the range [$-W/2,W/2$] such that $W$ measures the degree of randomness. Hereafter, we vary $L_\perp$ to model the $\mathrm{Cu_{3}BHT}$ samples of different thicknesses. 
\par

Numerically, we calculate the participation ratio defined as $p_2$=1/$\sum\limits_i|\varphi_E(i)|^4$ with $\varphi_E(i)$ being the normalized wave function amplitude of the eigenket of energy $E$ on site $i$. For a system of length $L$, the participation ratio scales with $L$ as $p_2$=$L^{\tilde d}$ with $\tilde d$ being the spatial dimension for extended states and 0 for localized states, respectively. Then, for a system of size $L_\parallel \times L_\parallel \times L_\perp$, we expect the following scaling function if there exists an ALT\cite{Pixley_Andersonlocalization_2015,Wang_Berezinskii_2024}
\begin{equation}
\begin{gathered}
p_2{(E,W)}={L_{\parallel}^D}{f(L_\parallel/\xi)}+\phi{L_{\parallel}^{-y}}{\tilde{f}(L_\parallel/\xi)}
\end{gathered}\label{eq_2_3}
\end{equation}
\par
Here, $D$ is the fractal dimension of the wave function of the critical state. $\xi$ is the correlation length and diverges at the critical disorder as $\xi\propto{|W-W_c|}^{-\nu}$. $\nu$ is the critical exponent characterizing the universality of the ALT. $\phi$ is a constant and $y$ is the exponent for the irrelevant variable. $f$ and $\tilde{f}$ are the scaling functions for the relevant and irrelevant variables that cannot be ignored for systems of relatively small sizes. 
\par

We fit our numerical data with the above scaling function and calculate the following quantity $Y_{L_\parallel}{(E,W)}={L_{\parallel}^{-D}}{[p_2{(E,W)}-\phi{L_{\parallel}^{-y}}{\tilde{f}(L_\parallel/\xi)]}}$. The presence of an ALT is judged by the following criteria: (i) For extended (localized) states, $Y_{L_\parallel} (E,W)$ increases (decreases) with the system size $L_\parallel$; (ii) At the critical point of the ALT, $Y_{L_\parallel} (E,W)$ is size-independent, and data near the critical point merge to a smooth scaling function $f(x)$. 
\par

We first consider the two-dimensional case. Since our model is spinless, we expect that all states are localized and there is no ALT. The $Y_{L_\parallel}{(E,W)}$ as a function of disorder $W$ at $E$=0 are shown in Fig. 4a for $L_\perp$=1. Clearly, $Y_{L_\parallel}{(W)}$ decreases with  $L_\parallel$ for even very weak disorders, a typical feature of localized states. This result is consistent with the well-established one-parameter scaling theory and validates the correctness of our scaling function. 
\par

Then, we consider a quasi-three-dimensional by setting $L_\perp$=16 and $L_\parallel$=40, 50, ..., 80 such that the ratio $L_\parallel/L_\perp$ is close to those in our experiments. Different from the two-dimensional case,  $Y_{L_\parallel}{(W)}$ crosses at a single point $W_c=10.7\pm$0.9 and data near the critical disorder merge to the scaling function with two branches representing extended and localized states; see Figs. 4b and c. This naturally explains the ALTs observed in our 3D samples with thicknesses ranging from 600-1200 nm. 
\par
We also perform the finite-size scaling analysis for samples with relatively small thicknesses. A representative result is shown in Fig. 4d for $L_\parallel$=3. We seem to have observed a crossing point representing the critical point of an ALT. However, such a crossing point is very close to 0, similar to the two-dimensional case. Therefore, we expect the thickness of the $\mathrm{Cu_{3}BHT}$ samples should be large enough for observing the ALT. This could help explain the lack of ALTs in our intermediate group as the dimensionality crossover drives the critical disorder down to 0 when the thickness of the sample decreases.
\par\quad\par

\noindent{\large\textbf{Discussions}}\\
The experimental results demonstrate both disorder-driven and dimensionality-driven transitions from metallic to insulating states in a benzenehexathiol-coordinated copper metal-organic framework. Given the separation between the voltage probes in our devices was sub-10 um, the ratio of longitudinal transport distance to the thickness of our films was around 100 in the thinner samples, 30 in the intermediate group, and 10 in the thickest samples, consistent with the range considered in our model.
\par

The temperature dependence of samples with different thicknesses illustrates the distinct behavior between metallic and insulating regimes of electronic transport. Furthermore, a detailed analysis of the scaling of conductivity with temperature supports the crossover from 3D to 2D transport behavior, based on Mott’s VRH (variable range hopping) theory.
\par

The disorder in these thin films mainly originates from domain boundaries as indicated by AFM and TEM images. The specifics of the disorder may vary from sample to sample as shown in Fig. S4, but the overall types and distribution remain consistent from our method of synthesis. We attribute the increased surface roughness in thicker films as shown in the topography data to the thickening process, which supports a higher overall disorder in our 3D samples compared to the quasi-2D ones. When we scrutinized the domain sizes between films of different thicknesses, we found no qualitative difference as the crystal domains' lateral length scale remains on the order of 100 nm (Fig. S5). This, along with the fact we observed metallic behavior in the thicker films supports that stochastic variations in the specific disorder arrangement do not jeopardize our statistical analysis of the metal-insulator transitions.
\par

Lastly, we note that $\mathrm{Cu_{3}BHT}$ has been previously reported as a superconductor with a transition temperature of 300 mK\cite{huang_superconductivity_2018}, indicating these transitions could evolve at lower temperatures into a superconductor-insulator transition (SIT) or superconductor-metal transition (SMT)\cite{Pan_BHTSC_2024}. In the current study, we found no sign of any global superconducting phase in the samples down to 1.8 K, so we stipulate the metal-insulator transitions as we observed. It is, however, possible that the same kind of disorder and dimensionality crossover would result in anomalous quantum criticalities near and below the superconducting transition temperature, which would be a fascinating topic in itself to study in future works.
\par

In conclusion, we have conducted a systematic quantum-transport study in mesoscopic devices made with $\mathrm{Cu_{3}BHT}$ thin films across a range of thicknesses. Our fitting results from the temperature dependence of the longitudinal conductivity reveal a dimensionality crossover from three dimensions to two. We found metal-insulator transitions both among 3D samples and between 3D and quasi-2D samples. In some intermediate samples close to the dimensionality crossover, we observed a sharp dip in the magnetoresistance data, which may be attributed to partial superconducting of the sample. Our theoretical modeling within the framework of Anderson localization transitions agrees with our experimental results and captures the localization physics behind these metal-insulator transitions. Moreover, it establishes for the first time that organic crystals such as conductive MOFs can exhibit quantum interference effects at low temperatures. With the versatile chemical designs and rich crystalline structures that come with organic materials, our results mark a milestone for liquid-liquid interface synthesized MOFs and open whole new avenues to search for novel quantum phenomena in organic material platforms.
\par\quad\par
\par\quad\par

\noindent{\large\textbf{Methods}}\\
\emph{Preparation of $Cu_{3}BHT$}
\par 
$\mathrm{Cu_{3}BHT}$ crystal was obtained via a reaction between $\mathrm{Cu^{2+}}$ ion and benzenehexathiol (BHT) molecule at the chloroform-water interface. BHT powder ($98\%$, Bide Pharmatech Co.,Ltd) was dissolved in degassed chloroform (HPLC, purity $\ge$ 99.9 wt.$\%$, Tianjin Jiangtian Chemical Technology Co., Ltd.) with a concentration of 1 mM under argon atmosphere. Then the solution was poured into a flask and kept at a temperature of 45 ℃. Preheated water with the same temperature was slowly added to the flask to form a stable oil-water interface. Then different volume of  1 M $\mathrm{Cu(OAc)_2}$ ($98\%$, Bide Pharmatech Co.,Ltd) aqueous solution was slowly added into the water phase until reaching particular concentration between 0.1 mM and 8 mM for sample thicknesses ranging from 10 nm to 800 nm. The Cu/BHT ratios in different groups are kept at 3. After 24h, the reaction was finished and the solution was gently removed. $\mathrm{Cu_{3}BHT}$ crystals were rinsed with degassed water, dimethyl formamide, and ethanol for at least three times each.
\par\quad\par

\noindent{\emph{Sample characterization}}
\par 
FTIR spectrums were obtained from Bruker Vertex 70 using attenuated total reflectance (ATR) mode. Thermo Fisher Scientific ESCALAB 250Xi was employed for XPS data. TEM, HRTEM, EDS and SAED results were obtained via Thermo Scientific Talos F200X. Surface morphology, surface potential, and local vertical conductivity were measured using Bruker Dimension Icon with tapping mode, KPFM mode, and TUNA module. Powder XRD data were obtained using a Rigaku SmartLab X-ray diffractometer with monochromatic Cu K$\alpha$ ($\lambda$ = 1.541 $\mathrm{\AA}$) radiation. The vertical and lateral conductivity of the $\mathrm{Cu_{3}BHT}$ film was taken using a Keithley 4200 semiconductor parameter analyzer. Detailed fabrication process are described in the device fabrication part. The contact resistance in the 2-probe measurement of vertical resistance was manually removed when calculating the vertical conductance.
\par\quad\par

\noindent{\emph{Device fabrication}}
\par 
To measure the vertical conductivity of $\mathrm{Cu_{3}BHT}$, we fabricated sandwich-like devices shown in Fig. S10. The bottom electrodes with 5 nm-thick chromium and 25 nm-thick gold were thermally evaporated onto the $\mathrm{SiO_2/Si}$ substrate using shadow mask. After $\mathrm{Cu_{3}BHT}$ crystals were transferred onto the electrodes, 30 nm-thick gold electrodes were thermally evaporated as top electrodes. The structure contains an effective overlapping area of 50 $\mu$m × 50 $\mu$m.
\par 

The four-probe devices and Hall-bar devices of $\mathrm{Cu_{3}BHT}$ were constructed with bottom-contact configurations since the $\mathrm{Cu_{3}BHT}$ films are not fully compatible with lithographic processes. 
\par 

For four-probe devices, four parallel 30 nm-thick gold electrodes with width of 50 $\mu$m and intervals of 100 $\mu$m were thermally evaporated onto the $\mathrm{SiO_2/Si}$ substrate using shadow mask. The rinsed $\mathrm{Cu_{3}BHT}$ thin films were suspended in degassed IPA solution and transferred onto the electrodes. Then the devices were annealed at 80 ℃ in a vacuum oven for 30 mins. 
\par 

For Hall-bar devices, 4 nm-thick chromium adhesion layer and 15 nm-thick gold electrodes were pre-patterned using photolithography and sputter deposited on $\mathrm{SiO_2/Si}$ substrates. The transfer method and annealing process is  same with four-probe devices.
\par\quad\par

\noindent{\emph{Transport measurements}}
\par 
Transport measurements were carried out in Physical Property Measurement Systems with a 9-Tesla magnet (Dynacool, Quantum Design). Two lock-in amplifiers (SR830, Stanford Research) were used to measure the $R_{xx}$ and $R_{xy}$ separately for magnetoresistance and Hall-effect measurements. All data were obtained with a maximum current of 1 $\mu$A to avoid sample heating by the Joule effect.
\par\quad\par

\clearpage
\bibliographystyle{sn-nature}

\clearpage

\noindent{\large\textbf{Acknowledgements}}\\
The authors acknowledge financial support from the National Key R\&D Program (2022YFB3603800, 2022YFA1204401, 2021YFB3600700), the National Natural Science Foundation of China (52373250, 52003190, 52121002, U21A6002).
\par\quad\par

\noindent{\large\textbf{Author contributions}}\\
J. Cheng, H. Ren, and W. Hu conceived the idea. J. Cheng prepared the samples and characterized the material. J. Cheng, S. Ding, and H. Ren conducted the transport measurements. J. Cheng, S. Ding, F. Yang, and H. Ren analyzed the data. W. He and J. Wang contributed to the fabrication of Hall bar devices. Y. Pang. helped with the sample characterization. C. Wang performed the theoretical calculations. S. Ding, H. Ren, and W. Hu supervised the study, interpreted the results, and revised the paper. All authors prepared the manuscript. 
\par\quad\par

\noindent{\large\textbf{Competing interests}}\\
The authors declare no competing interests.
\par\quad\par

\newpage

\begin{figure}[htbp]
\centering
\includegraphics[width=1.0\textwidth]{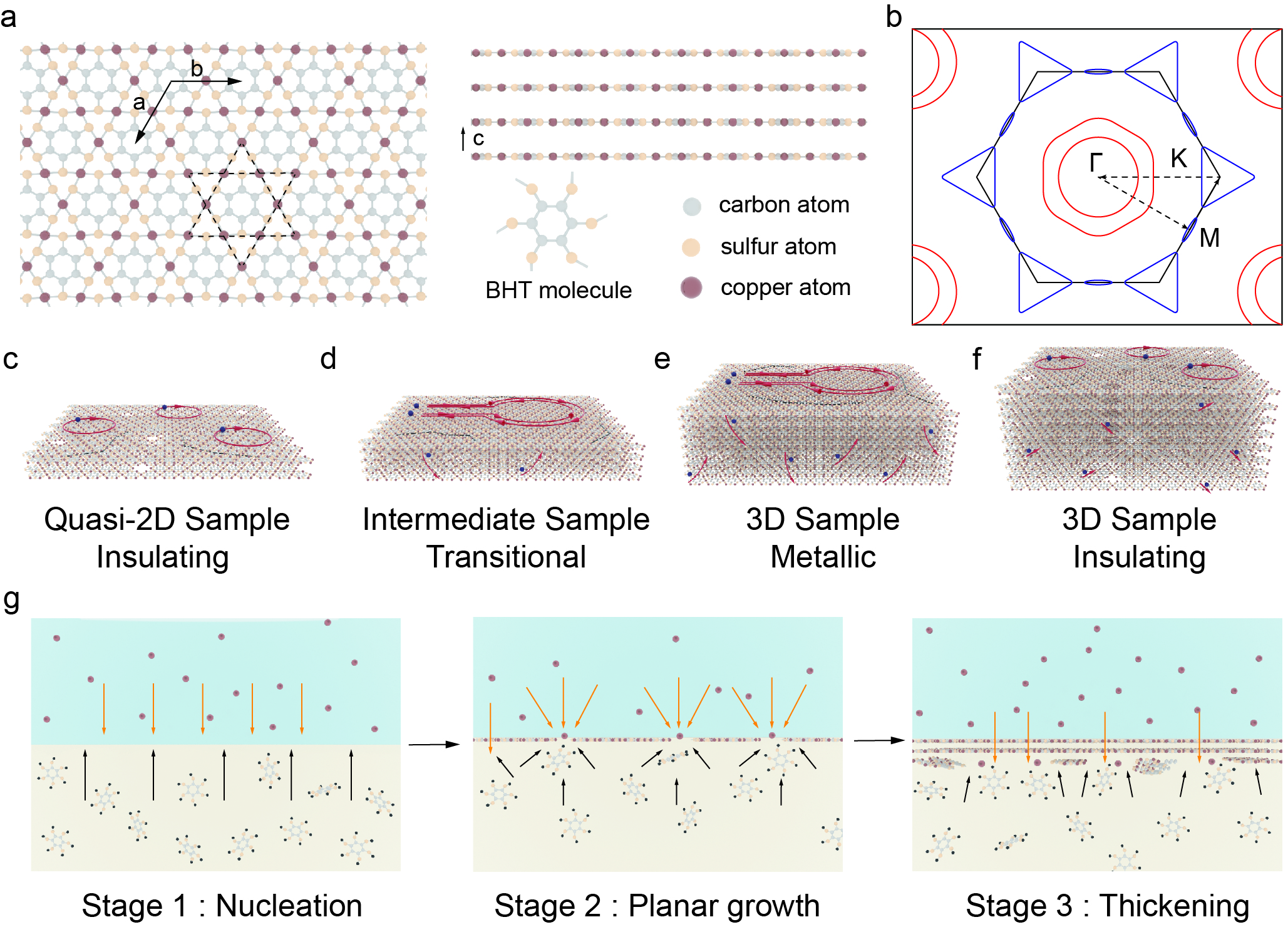}
\caption{\textbf{Schematic diagram of the structure, metal-insulator-transition and synthesis process of} \textbf{$\mathrm{Cu_{3}BHT}$}. \textbf{a} Two-dimensional lattice and layer structure of $\mathrm{Cu_{3}BHT}$. \textbf{b} Fermi surface contour of $\mathrm{Cu_{3}BHT}$. \textbf{c} Localized electrons in quasi-2D samples with disorders. \textbf{d} Weakly localized electrons in intermediate samples with disorders. \textbf{e} Localized and extended states in 3D samples with disorders. \textbf{f} Localized states in 3D samples with increased disorders. \textbf{g} Different stage of liquid-liquid interface synthesis of $\mathrm{Cu_{3}BHT}$.}
\label{fig_1}
\end{figure}

\newpage

\begin{figure}[htbp]
\centering
\includegraphics[width=1.0\textwidth]{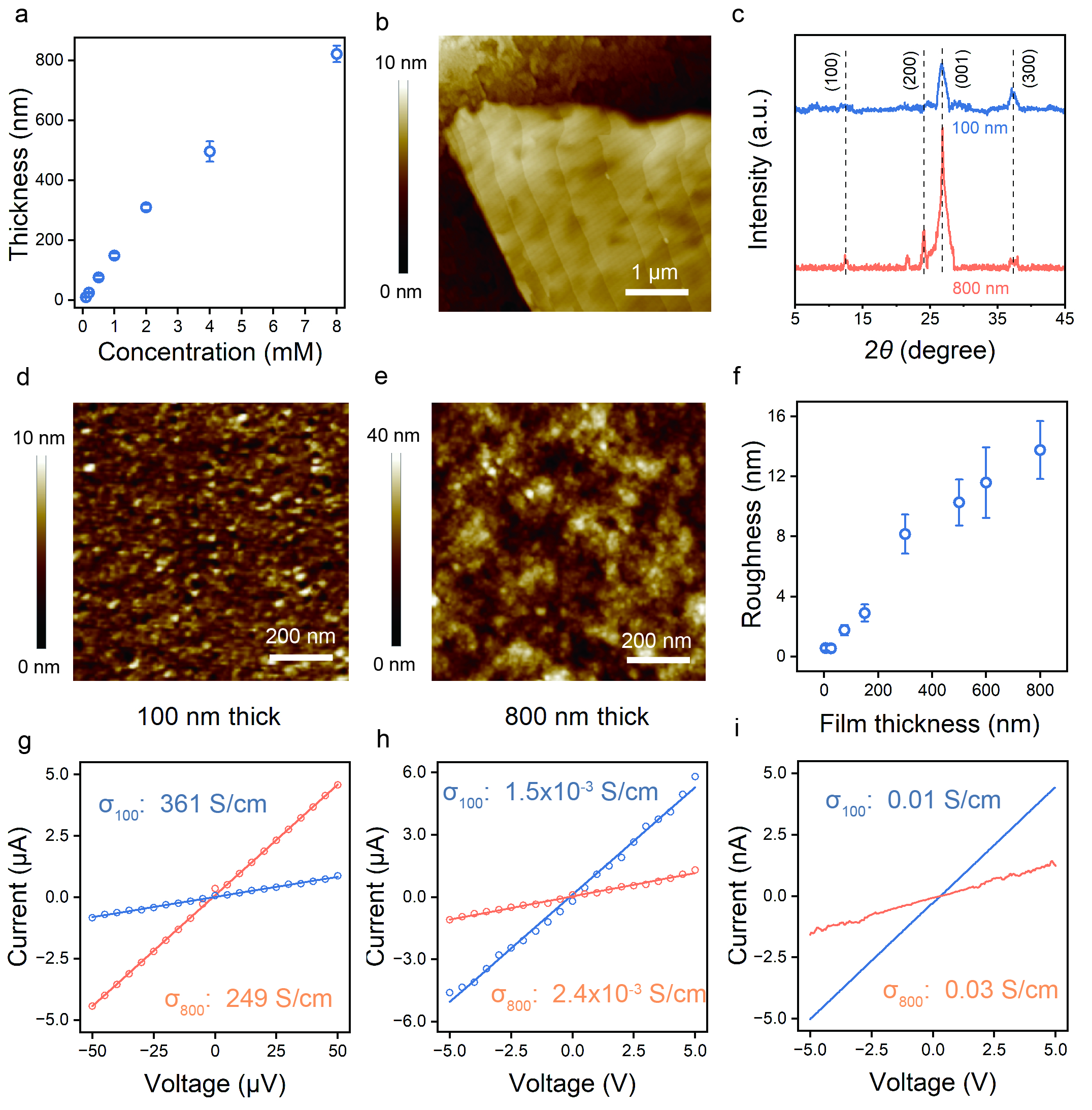}
\caption{\textbf{Room temperature characterization of $\mathrm{Cu_{3}BHT}$ with different thicknesses.} \textbf{a} Relationship between $ \mathrm{Cu^{2+}}$ concentration and film thickness. \textbf{b} AFM image of layered structure of $\mathrm{Cu_{3}BHT}$ film. \textbf{c} PXRD pattern of 100 nm- and 800-nm thick $\mathrm{Cu_{3}BHT}$. \textbf{d} Surface roughness of 100 nm-thick $\mathrm{Cu_{3}BHT}$ film. \textbf{e} Surface roughness of 800 nm-thick $\mathrm{Cu_{3}BHT}$ film. \textbf{f} Relationship between film thickness and surface roughness. \textbf{g} $I$-$V$ curve and estimated conductivity on the lateral direction of 100 nm- and 800 nm-thick $\mathrm{Cu_{3}BHT}$ samples measured by probe station. \textbf{h} $I$-$V$ curve and estimated conductivity on the vertical direction of 100 nm- and 800 nm-thick $\mathrm{Cu_{3}BHT}$ samples measured by probe station. \textbf{i} $I$-$V$ curve and estimated conductivity on the vertical direction of 100 nm- and 800 nm-thick $\mathrm{Cu_{3}BHT}$ samples measured by Peakforce TUNA module.}
\label{fig_2}
\end{figure}

\newpage

\begin{figure}[htbp]
\centering
\includegraphics[width=1.0\textwidth]{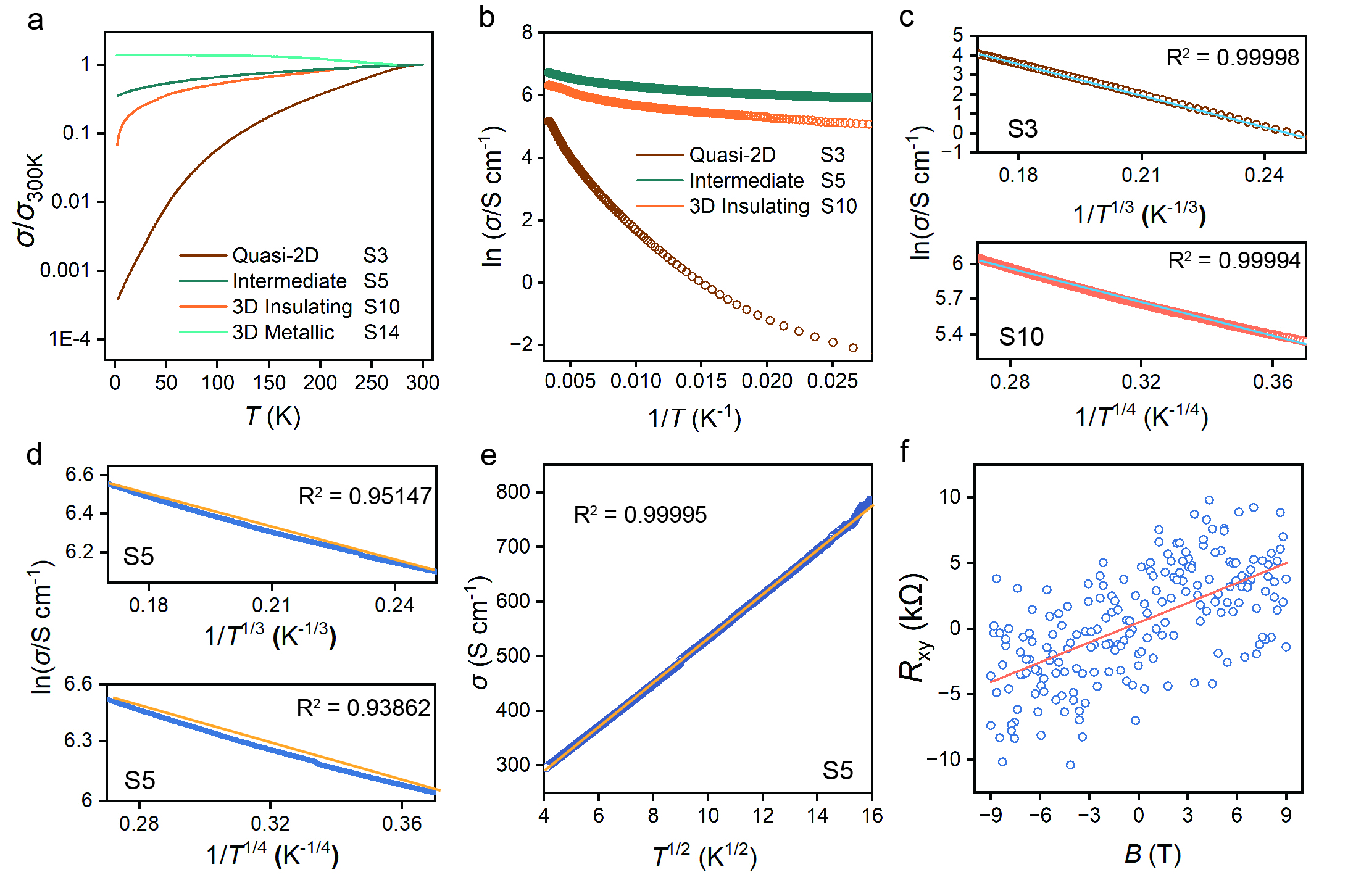}
\caption{\textbf{Metal-insulator transitions and Hall measurement of $\mathrm{Cu_{3}BHT}$.} \textbf{a} Normalized conductivity-temperature curve of samples in different groups. \textbf{b} Plots of ln($\sigma \mathrm{/S \; cm^{-1}}$) versus {1/$T$} of the insulating samples S3, S5 and S10. \textbf{c} VRH fitting of the insulating samples S3 and S10. \textbf{d} VRH fitting of sample S5 in the intermediate group. \textbf{e} $\sigma \sim T^{\frac 1 2}$ curves of sample S5 in the intermediate group. \textbf{f} Transverse resistance of sample S8 at 5K.}
\label{fig_3}
\end{figure}

\newpage

\begin{figure}[htbp]
\centering
\includegraphics[width=1.0\textwidth]{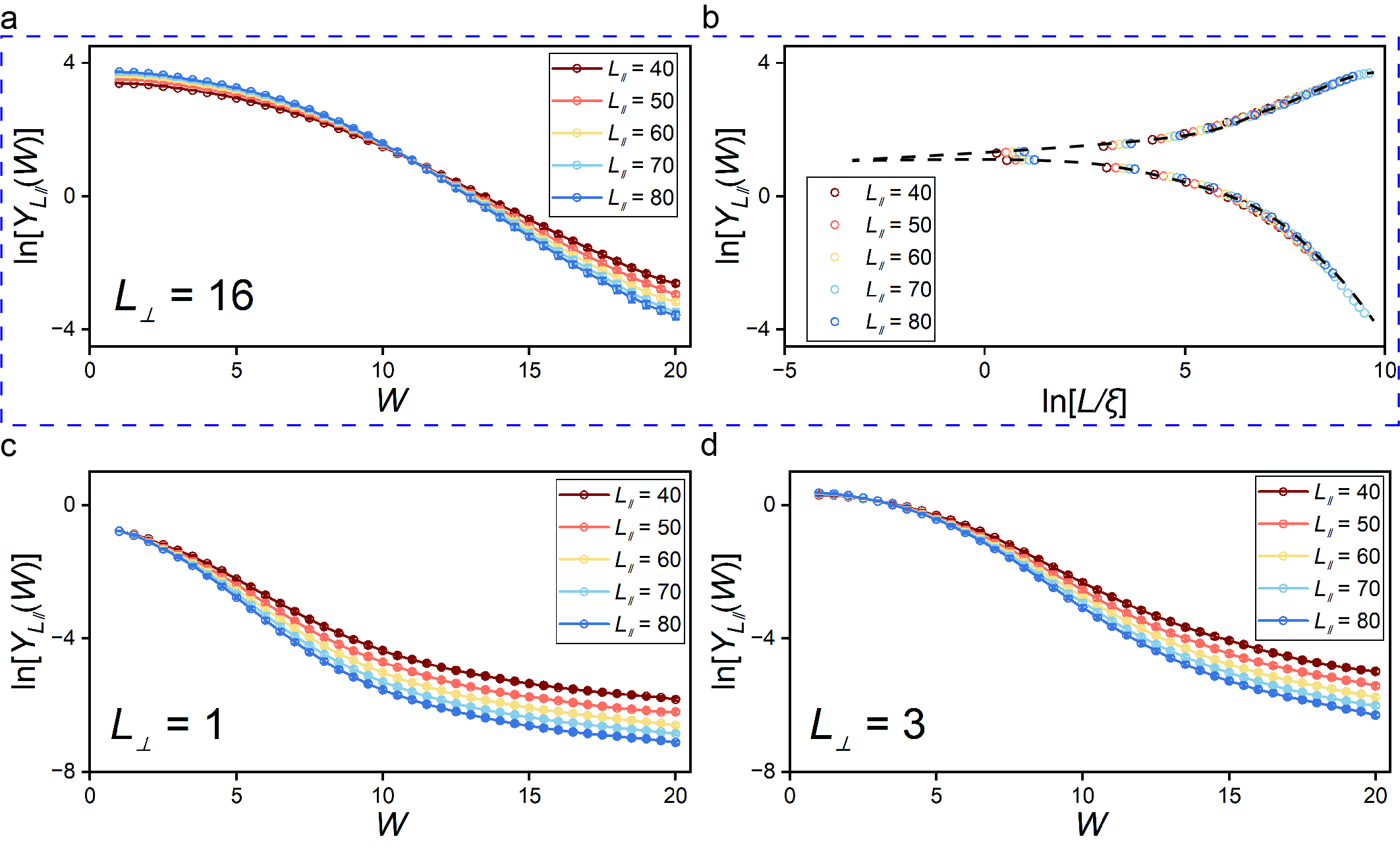}
\caption{\textbf{Theoretical modeling of Anderson localization transitions} \textbf{a} Scaling variable ln$Y_{L_\parallel}$  as a function of disorder strength W for $L_\perp$=16  and different $L_\parallel$ = {40, 50,... 80} . \textbf{b} Scaling function ln$[Y_{L_\parallel}]$ =ln$[f(\mathrm{ln}[l/\xi])]$ obtained by merging data near the crossing point in \textbf{a} into two branches of curves (upper and lower stand for extend and localized states respectively). \textbf{c} Same as \textbf{a} but for $L_\perp=1$  where no ALT is seen. \textbf{d} Same as \textbf{a} but for $L_\perp=3$, an intermediate height between those that can and cannot support the ALT.}
\label{fig_4}
\end{figure}

\end{document}